# Summary overview of present state of basic electrostatic field electron emission theory

Richard G. Forbes

University of Surrey, School of Mathematics and Physics, Guildford, Surrey GU2 7XH, UK

Electronic mail: r.forbes@trinity.cantab.net

This technical note provides a high-level overview of the present state of basic field electron emission (FE) theory, as suitable for use in the context of technological applications of FE theory. At present there is much theoretical confusion in FE literature, and a partial breakdown of the peer review system. Even in sensitive technological contexts, many papers have stated and used out-of-date theory that makes current-density predictions that are several hundred times less than those of modern FE theory. A primary aim of this note is to help reduce the confusion and error in future published FE literature. It is not intended as a detailed review of FE theory.

## I. THEORETICAL BASICS

Theoretical aspects of the published literature of field electron emission (FE) can be confusing or misleading. This document aims to provide a brief overview of the present state of basic electrostatic FE theory, as used in technological contexts, at a level that (hopefully!) can be easily understood by non-experts. Notation here is more precise than is commonly found in past literature. Note that the values of universal constants are given to seven significant figures, but (obviously) calculation results should be appropriately rounded.

The term "electrostatic field electron emission (ESFE)" is being used here because papers are being published that assume that the electric component of a travelling electromagnetic (EM) wave can also induced tunnelling-type emission. At the present time, it is not clear whether this "electromagnetically-induced FE (EMFE)" is physically possible or not, because EM radiation normally interacts with matter via photon absorption. The present overview deals only with the theory of FE induced by a constant electrostatic field.

Although there are many more-sophisticated theoretical treatments, most "technological-level" treatments of electrostatic field electron emission are based on a Sommerfeld-type free-electron model that assumes the emitter surface is smooth and planar. The atomic structure of the emitter and detailed band-structure effects are disregarded. Various different assumptions can then be made about the nature of the potential-energy (PE) barrier encountered by an escaping electron and about the quantum-mechanical formalism used to evaluate transmission probability. As a result, there are many different "smooth-planar ESFE equations", corresponding to different levels of approximation. The two most commonly used equations can be called the "elementary FE equation" and the "1956 Murphy-Good (zero-temperature) FE equation" [this is the zero-K version of a finite-temperature formula derived by Murphy and Good].

Although this type of theory is derived for flat surfaces, it can also be applied "adequately" to curved surfaces, provided the curvature is "not too great". In particular, it can be applied to needle-shaped emitters and to protrusion-type emitters on surfaces, provided that the emitter is "not too sharp" (sometimes interpreted as having an apex radius of curvature greater than 20–50 nm).

## II. THE TWO COMMON FE EQUATION FORMS

The *elementary (EL) FE equation* is a simplified version (apparently introduced in 1978 [1]) of the various FE equations that were discussed in the 1920s. The elementary FE equation and these 1920s equations are all based on the theoretical assumption that tunnelling can be satisfactorily modelled as taking place through an exactly triangular (ET) PE barrier. The elementary FE equation gives the local emission current density $J$, in terms of the local work function $\phi$ and the magnitude $F$ of the (negative) local surface ES field, as

$$J \sim J^{\mathrm{EL}} \equiv a\phi^{-1}F^2 \exp[-b\phi^{3/2}/F] , \qquad (1)$$

where $a$ [$\cong 1.541434$ μA eV V$^{-2}$] and $b$ [$\cong 6.830890$ eV$^{-3/2}$ V nm$^{-1}$] are the first and second Fowler-Nordheim (FN) universal constants, respectively. This equation is often called the "Fowler-Nordheim equation", but does not in fact appear in the 1928 FN paper or in other 1920s FE literature.

The *1956 Murphy-Good (MG) (zero-temperature) FE equation* is based on the theoretical assumption that tunnelling can be satisfactorily modelled as taking place through the so-called *Schottky-Nordheim (SN)* barrier (see Appendix). This barrier and equation take exchange-and-correlation effects into account, by approximating these effects by an image potential energy. The resulting equation has the overall form

$$J \sim J^{\mathrm{MG0}} \equiv \lambda_{\mathrm{p}}(\phi,F) \cdot a\phi^{-1}F^2 \exp[-v_{\mathrm{e}}(\phi,F) \cdot b\phi^{3/2}/F] , \qquad (2)$$

where $\lambda_p(\phi,F)$ and $\nu_e(\phi,F)$ ($\nu$ is "nu") are well-defined "pre-exponential" and "exponent" correction factors that nowadays can be evaluated precisely to any desired degree of precision.

Both factors can be expressed as functions of a single modelling variable, which can be either the Nordheim parameter $y_F$ or a parameter $f$ now called the "scaled field for an SN barrier of zero-field height $\phi$". The relationship between these parameters is: $f=y_F^2$. The $y_F$-based approach derives from the 1953 work of Burgess, Kroemer and Houston (BKH) [2]; the $f$-based approach derives from work by Forbes and Deane (FD) in the period 2006 to 2010 (see [3] for a recent overview). Both approaches yield the same numerical results, and both approaches are currently acceptable, but the FD approach is considered mathematically superior and better suited to further mathematical development of FE theory and data analysis. Most papers that discuss Murphy-Good FE theory still use the BKH approach, but the FD approach is used here.

The field-magnitude needed to reduce a SN barrier of zero-field height equal to the local work function $\phi$ is called the *reference field*, is denoted by $F_R$ and is given by

$$F_R \;=\; c^{-2}\phi^2, \qquad (3)$$

where $c\;[\equiv (e^3/4\pi\varepsilon_0)^{1/2}]$ is the *Schottky constant* (sometimes denoted by $c_S$). Here, $e$ is the elementary (positive) charge and $\varepsilon_0$ is the electric constant (aka the permittivity of free space). The *scaled field* $f$ (for an SN barrier of zero-field height $\phi$) is then given by

$$f \;\equiv\; F/F_R \;=\; c^2\phi^{-2}F \;=\; (e^3/4\pi\varepsilon_0)\phi^{-2}F \;\approx\; 1.439965 \cdot (\mathrm{eV}/\phi)^2 \cdot \{F/(\mathrm{V\ nm}^{-1})\}\ . \qquad (4)$$

In the FD approach, the correction factor $\nu_e(\phi,F)$ in eq. (2) is given by $v_{FD}(f)$ and the correction factor $\lambda_p(\phi,F)$ is given by the expression $\{t_{FD}(f)\}^{-2}$, where $v_{FD}(f)$ and $t_{FD}(f)$ are particular applications of special mathematical functions $v_{FD}(x)$ and $t_{FD}(x)$ introduced and defined by FD. The parameter $x$ is an abstract mathematical variable called by FD the *Gauss variable* (because it can be identified with the independent variable in the Euler/Gauss Hypergeometric Differential Equation). Equation (2) then takes the form:

$$J \sim J^{MG0} \equiv \{t_{FD}(f)\}^{-2} \cdot a\phi^{-1}F^2 \exp[-v_{FD}(f) \cdot b\phi^{3/2}/F] . \qquad (5)$$

Exact mathematical procedures for precisely specifying and evaluating $v_{FD}(f)$ exist. However, in most technological contexts, an adequate approximation for $v_{FD}(f)$ is

$$v_{FD}(f) \approx 1 - f + (1/6) f \ln f . \qquad (6)$$

Over the range $0 \leq f \leq 1$, the value of $v_{FD}(f)$ diminishes monotonically from 1 to 0. Over the whole of this range, the accuracy of approximation (6) is better than 0.33%. For a 4.5 eV work-function emitter, $f$ typically lies within the range $0.15 < f < 0.35$, but can be higher than this.

An exact mathematical procedure does exist for evaluating $t_{FD}(f)$, but it is usually sufficient to approximate $\{t_{FD}(f)\}^{-2} \approx 0.9$. [Or to approximate it as 1.0 and leave it out of the equation.]

## III. WHICH EQUATION IS BETTER?

The Murphy-Good FE formula, eq. (2) or (5), predicts local-emission-current-density values that are much greater than those given by the elementary FE equation, eq. (1), typically by a factor of several hundreds. Thus, there is an issue as to which equation is better.

Large parts of modern chemistry, modern condensed matter physics and modern density-functional theory assume that exchange-and-correlation (E&C) effects are important in modern science. It is not logically tenable to assume that they will not be important at metal surfaces. Thus, it has long been theoretically certain that Murphy-Good FE theory (which, as noted above, models E&C effects by an image potential energy) is better physics than either 1920s-type FE equations or the elementary FE equation.

This issue has previously been discussed in Ref. [4].

The superiority of Murphy-Good FE theory over earlier FE theory has been pointed out in a large number of FE research textbooks and review articles published since 1960. A list of 25 relevant textbooks and review-type articles can be found by a websearch on the title <Support for Murphy-Good FE theory in textbooks, reviews, etc.>.

You can also find relevant information by typing the question: "What is the correct modern theory of field electron emission?" [or the question "What is Murphy-Good field emission theory"] into a web browser, and then exploring the information given by the Google AI assistant. My experience is that you get different detailed answers, depending on precisely how you formulate the question and any follow-up requests. However, the AI assistant usually gets the basic point about the superiority of Murphy-Good theory correct, although it is not always fully correct about mathematical or other details.

Use of eq. (1) is still quite common in FE literature, despite the facts that clearly eq. (1) is significantly worse physics, and clearly eq. (1) is much worse numerically, than eq. (2). These things have been known to theoreticians since the late 1950s. What this situation shows is that the peer review system in FE has to some extent broken down. This breakdown seems one of the biggest problems in the subject area. Further confusion is caused by authors who refer to eq. (2) as the "Fowler-Nordheim equation". [Others then assume that the term "Fowler-Nordheim equation" refers to an equation in the 1928 FN paper, or to the elementary FE equation.] At present, there is no consensus about which particular equation should be called the "Fowler-Nordheim equation": in my view, best policy would be that this simple name should not be used in future: different FE equations should each be given their own names, as is done here.

Overall, the literature of FE can be very confusing, particularly for non-experts, and it is not surprising that many authors seem to have had difficulty finding out what "good modern understanding" of FE actually is. Hopefully this situation should get better as people realize that the Google AI assistant is a helpful tool for showing where to look for modern FE theory (even if the tool is not always correct in detail).

## IV. MURPHY-GOOD FE THEORY AS TRANSITIONAL

Note that Murphy-Good FE theory, even in its modern "21$^{st}$ Century" form, is a *transitional* theory. More-sophisticated theories will soon be needed that take into account: (a) the detailed effects of atomic structure and band structure; and/or (b) the behavior of sharply curved emitters; and/or (c) deep issues relating to the quantum mechanics of tunnelling. However, it seems to me that (in many or most cases) so-called "Extended Murphy-Good FE theory" is "good enough as a working theory for the time being". There are advantages if everybody uses *the same* theory to interpret their experimental data, even if the extracted characterization parameters are not precisely correct.

Advances have been made in developing more-sophisticated FE theory—for example the theory by Kyritsakis and colleagues relating to FE from an Earthed spherical emitter [5, 6], but mostly the situation is one of "theoretical research in progress". It is outside the scope of this document to review advanced treatments of FE theory.

## V.  FACTORS OTHER THAN EMISSION PHYSICS

Note that this overview is about the *emission physics* of FE. In technological contexts, an emitter is always part of an electrical engineering FE system, and an associated electrical circuit. Thus, the measured current-voltage $I_m(V_m)$ characteristics may be partially determined by factors other than the emission physics (see [7]). An FE system where the emission physics is effectively the exclusive influence on the $I_m(V_m)$ characteristics has been termed *electronically ideal*. For emitters that are "not too sharp", there exist *validity tests* that can be applied to measured current-voltage data, to check whether an FE system is electronically ideal (or nearly so) (e.g., see [8]).

Some causes of non-ideality, such as field emitted vacuum space-charge, have been well researched. But there is a need for more research on other possible causes of non-ideality [7], in particular "voltage loss" along an emitting needle.

Even for electronically ideal systems where the emitter is "not too sharp", accurate extraction of emitter characterization data (particularly emission-area information) by using the well-known Fowler-Nordheim plot [$\ln\{I_m/V_m^2\}$ vs $1/V_m$] can be messy, because Murphy-Good FE theory shows that a FN plot is theoretically expected to be slightly curved, particularly for lower values of $1/V_m$. (For a recent discussion, see [9]). A better alternative, namely the *Murphy-Good plot*, has been suggested [10,11], but is not in wide use. (As shown clearly in [11], Murphy-Good plots are theoretically expected to be significantly "more nearly linear" than Fowler-Nordheim plots, and hence are easier to interpret)

# VI. THE NEED FOR SCIENTIFIC COMPARISONS

To move on decisively, it would be helpful to have good scientific comparisons between FE theory and experiment, for a well-defined FE system that is electronically ideal (or nearly so).

In the last 100 years there have been thousands of experiments on the electrostatics of field emitters. However, there has been only **one** known published experiment [1], reported in 1978, that has attempted to test FE theory quantitatively and (more-or-less) reliably against experiment. This experiment appears to suggest that our best guess should be that, for planar atomically structured surfaces, the "true predicted local emission current density" will lie between the predictions of eqns (1) and (2), but closer to eq. (2). A new form of experiment, for which a "proof of concept" has recently been submitted for publication [7], tends to confirm the earlier result.

# VII. FURTHER AND FUTURE INFORMATION

The best published account of the Forbes-Deane approach to FE theory used above is Ref. [3], but unfortunately the chapter concerned is behind a paywall, albeit a relatively small one. Several freely accessible tutorial lectures relating to FE theory can be found by a websearch on <ResearchGate Richard G. Forbes>. The more recent ones should be considered more useful.

References [4, 7, 8 10, 11] are open access. Free-to-download versions of [5] and [6] may be obtained from ResearchGate, via a websearch on the title. A pre-review version of [9] is available via the arXiv url shown.

A version of the present document is also available on ResearchGate, and may be updated from time to time. In future, the latest version of the document may be found by a websearch on its title (not including the date).

## APPENDIX: THE SCHOTTKY-NORDHEIM BARRIER

In a one-dimensional treatment of a planar emission situation, let $z$ denote distance outwards, normal to the planar surface. The quantity that determines electron transmission probability is the difference between the electron potential energy $U_e(z)$ and the normal component ($E_z$) of the electron total-energy. Both these quantities must be measured relative to the same energy reference zero. The author uses the symbol $M$ to denote this difference and calls it the *motive energy*. Thus, in general:

$$M(z) = U_e(z) - E_z. \qquad (7)$$

The Schottky-Nordheim (SN) barrier is the particular barrier defined by:

$$M^{SN}(H,F,z) = H - eFz - e^2/16\pi\varepsilon_0 z. \qquad (8)$$

Here, $H$ is the zero-field height of the barrier, and the other symbols have the same meanings as above. The final term in eq. (8) is the image potential-energy term. The SN barrier with zero-field height $H$ equal to the local work function $\phi$ is of special interest. The exactly triangular (ET) barrier, $M^{ET}(H,F,z)$ omits the image-PE term.

The two different barrier forms just discussed are illustrated in Fig. 1 (the vertical axis shows the motive energy in eV).

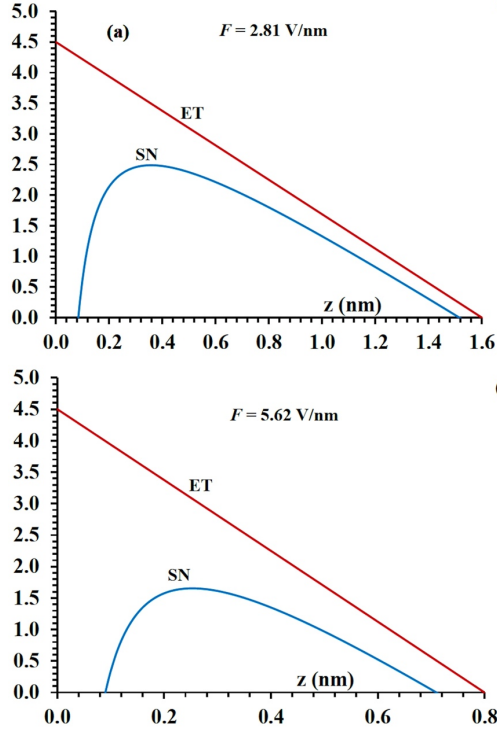

Fig. 1. To compare the forms of the Schottky-Nordheim (SN) and exactly triangular (ET) tunnelling barriers, for barrier height $H$ equal to a local work-function value of 4.50 eV, and for local field values equal to 2.81 V/nm ($f$=0.2) (upper diagram) and 5.62 V/nm ($f$=0.4) (lower diagram). (Diagrams prepared by Dr M.M. Allaham. Similar diagrams have been published in Ref. [11].)

## AUTHOR DECLARATIONS

**Conflicts of Interest**

The author has no conflicts to disclose.

## DATA AVAILABILITY

This technical note makes no use of experimental or theoretical data.